\documentclass[fleqn,10pt]{wlscirep}
\usepackage[utf8]{inputenc}
\usepackage[T1]{fontenc}
\usepackage{graphicx}
\usepackage{subcaption}
\title{Linear optical properties of a linear chain of interacting gold nanoparticles}

\author[1]{Asef Kheirandish}
\author[1,*]{Nasser Sepehri Javan}
\author[1]{Hosein Mohammadzadeh}
\affil[1]{Department of Physics, University of Mohaghegh Ardabili, P.O. Box 179, Ardabil, Iran}
\affil[*]{nsj108119@yahoo.com}


\begin{abstract}
In a Drude-like model for the conduction electrons of Metal Nanoparticles (MNPs) in a periodic linear chain, considering dipole-dipole interactions of adjacent particles, an analytical expression is derived for each particle permittivity for two different polarizations of incident light: parallel with and perpendicular to the chain line. A numerical analysis is carried out for a chain including 10 identical gold Nanoparticles (NPs) for two different sizes of NPs and two different host media of air and glass. It is shown that in the parallel case of polarization, interaction of NPs leads to a substantial increase in the extinction cross section and the red-shift of the Surface Plasmon Resonance (SPR) wavelength. In comparison with the linear properties of a single NP, the second and penultimate particles have the most increase in the extinction cross section and SPR wavelength displacement while the first and last particles experience the least variations due to the mutual interactions. For the perpendicular polarization, inversely, the dipolar coupling causes the decrease in extinction cross section of all particles and the blue-shift of SPR wavelength. For the parallel polarization, the absolute values of the real and imaginary parts of complex permittivity of each MNP decrease in comparison with the single particle case while they increase for the perpendicular state of polarization.
\end{abstract}
\begin{document}

\flushbottom
\maketitle

\thispagestyle{empty}

	\section*{\label{sec:level1}Introduction}

Theoretical and experimental studies related to the linear and nonlinear interaction of electromagnetic waves (EMWs) with MNPs are of great interest to science and technology. The main and characteristic effect occurring during EMWs interaction with media containing MNPs is the SPR origin of which is the collective oscillation of conduction electrons of MNPs. The SPR frequency region of noble MNPs is located at a visible spectrum \cite{kreibig2013optical} which makes them attractive for plasmonics and nano-photonics fields where beyond the diffraction limit, light can interact with structures whose dimensions are comparable or less than the incident light. Growing progress in nanotechnology has been made different possibilities to design aggregates and arrays of MNPs with different shapes and sizes which are periodic in one, two or three dimensions with feasibility of different applications including  nano-photonic ones such as optical waveguides \cite{quinten1998electromagnetic,brongersma2000electromagnetic,maier2005plasmonics,zhen2008collective,zakomirnyi2018titanium}, biosensors \cite{elghanian1997selective,haes2004nanoscale,zhang2018plasmonic}, subwavelength imaging \cite{simovski2005resonator,alitalo2006near}, optoelectronic devices \cite{maier2005plasmonics}, optical metamaterials \cite{pendry2003positively,pendry2006metamaterials}, sensors for detection of toxic heavy metals \cite{ding2016nanomaterial,chen2014colorimetric} and solar cells \cite{atwater2011plasmonics}.\\
In this theoretical study, we investigate the linear properties of identical MNPs placed in a one-dimensional linear periodic chain. In order to achieve analytical equations, for describing dynamics of conduction electrons of each MNP, we use classical motion equation including the dipole-dipole interaction of neighbor particles. Such a primitive model is commonly used for the cases in which the particles are not very close to each other and their characteristic dimension is very less than the incident light wavelength. For a very close spacing of particles and particle dimensions comparable with the wavelength, the multipolar interaction of particles and quantum tunneling of electrons become important. Coupled dipole approximation is a well-established approach in the nano-photonics. The scattering of light from a one-dimensional infinite periodic structure of NPs has been studied by Markel \cite{markel1993coupled,markel2019extinction}. The solutions for a finite chain has been also presented by Markel and coworkers \cite{markel2007propagation,rasskazov2014surface,rasskazov2014waveguiding}. Such a model has been employed in several theoretical studies regarding nano-waveguides, during propagation of light through a linear chain of MNP \cite{weber2004propagation,maier2003optical}. The ability of linearly ordered gold nanorods for advanced plasmonic applications has been experimentally studied by Kang et. al.\cite{kang2019dichroic} They used a simple method for fabrication of dichoric plasmon superstructures over a macroscopic area. The prepared system shows a polarization-dependent extinction which causes the selective excitation of transverse or perpendicular modes of SPR peaks. Nonlinear response of linear chain of atomic silver particles has been simulated in Ref\cite{yan2018plasmon}. They showed that the decay of hot electrons has a main role in the nonlinear dynamics of system. For such a linear chain of atomic silver particles, influence of nuclear dynamics on the temporal evolution of SPR has been theoretically investigated in Ref\cite{donati2018anisotropic}, and it has been shown that the anisotropic SPR transfer and eventual decay can be explained via non-radiative relaxation and changes in dipole-dipole polarizability during the nuclear dynamics.
The motivation of this work is finding analytical expressions for the permittivity of each MNP in a chain containing countable numbers of particles. Such an attempt for MNP dimer was successful in our recent work \cite{asef2020analytical} where findings were in the agreement with experimental data. In the linear regime, the measurable parameter which is a complicated function of the permittivity is the extinction cross section of particles which indicates the absorption and scattering of light, thus, we calculate it for each MNP of chain. Effects of geometrical and physical parameters such as surrounding medium, particles size and spacing and state of incident EMW polarization on the extinction cross section and complex permittivity of each MNP are considered. Its worth mentioning that in the dipole-dipole interaction approach for description of optical properties of MNP used by refs. \cite{markel2007propagation,rasskazov2014surface,rasskazov2014waveguiding}, interaction of all NPs is taken into account which is lead to some complicated expressions for main optical parameters. There, the essential parameter of model is the electric dipole moment of each MNP which is calculated through an expression inside of which exists the susceptibility of each individual particle that should be determined by the experiment or theory. Here, we consider only the interaction of two near neighbors of each NP in order to get some analytical relations for some crucial physical parameters. Inside the model, the permittivity of each interactional MNP is calculated analytically and is simplified under some realistic conditions. Obtaining analytical expressions for the permittivity is of great importance as it can show the role of different parameters in the linear response of correlated particles to the incident EMW. The analytical findings of this study can be very useful for future theoretical and experimental investigations. 

\section*{\label{sec:level2}Model}

According to Fig. (1), we consider the linear interaction of electric field of a normally incident EMW with a linear chain consisting of finite numbers of interacting MNPs.  There are two possible polarization states for the incident EMW (perpendicular to/parallel with the orientation of NP chain) which are demonstrated in Figs (1-a) and (1-b). The classical Thomson model \cite{thomson1904xxiv} is employed to describe the interaction of EMW with conduction electrons of spherical MNP. Despite being unsuccessful in atomic physics for describing the atoms behavior, however this model still is recognized as an efficient approach for classical theories of the light-cluster interaction phenomena\cite{kreibig2013optical,kheirandish2018polarization,asef2020analytical}. In this model, we suppose that the conduction electrons of ${N_c}$ individual metal atoms are homogeneously located inside the spherical volume of NP with radius $a$ and the background positive ions which are distributed homogeneously as well, are immobile. Considering the average separation of atoms as d, the atoms number density or equivalently the ions density is ${n_a} = 1/{d^3}$ while we use ${n_{0e}} = 3Z{N_c}e/(4\pi {a^3})$ for the density of conduction electrons in the equilibrium state, where $e$ is the magnitude of the electron charge and $Z$ is the contribution of each atom in the conduction electrons. In the model which is called rigid sphere where the particle radius is much less than the wavelength, $a << \lambda $, for the interaction of a small MNP with the low-intensity EM fields, the spatial variations of fields inside the MNP can be ignored. In this case, all conduction electrons experience the same forces at a moment and move identically. This situation is shown in Fig. (1-c), where the whole unified electronic cloud moves from its equilibrium state.\\
\begin{figure}
	\centerline{\includegraphics[scale=.40]{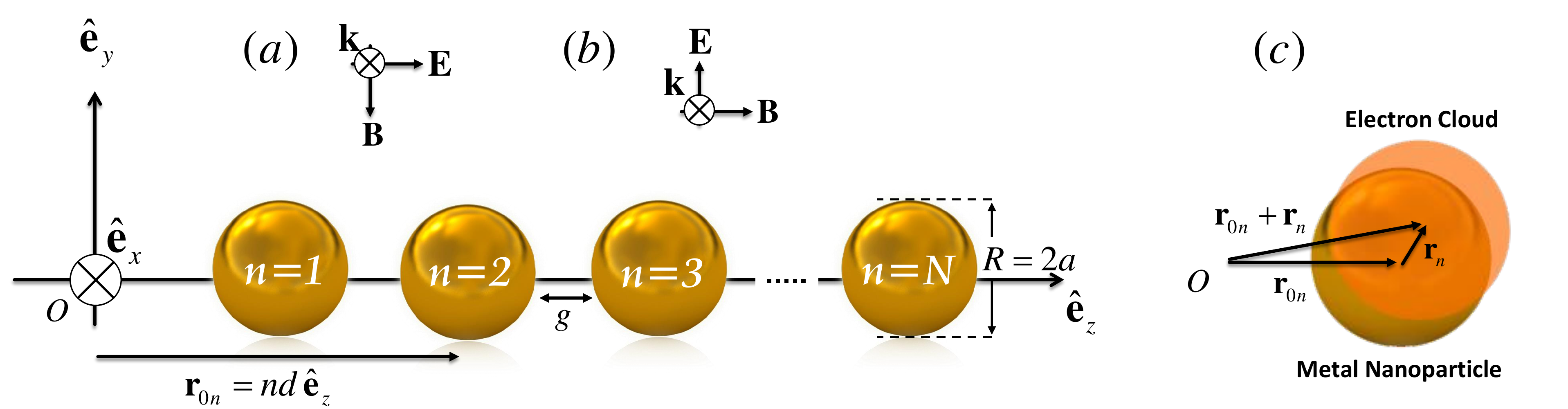}}
	\caption{Configuration of the problem for two different cases: (a) the laser electric field, (b) the laser magnetic field is parallel to the symmetry axis of the chain and (c) displacement of conduction electronic cloud from the initial equilibrium position.}
	\label{fig1}
\end{figure}
In the linear regime, taking into account the inter-particle interactions in the dipole-dipole approximation, the momentum equation of conduction electrons of $n$th MNP can be written as
\begin{align}
{m_e}\frac{{{d^2}{{\bf{r}}_n}}}{{d{t^2}}} + {m_e}\gamma \frac{{d{{\bf{r}}_n}}}{{dt}} + {m_e}\xi \omega _p^2{{\bf{r}}_n} =  - e{{\bf{E}}_n},
\end{align}
where ${{\bf{r}}_n}$, ${m_e}$, $\gamma $ and $\xi $ are the displacement of electronic cloud of $n$th MNP from the equilibrium state, the electron mass, damping factor and a function of MNP radius obtained from the experimental data \cite{asef2020modified}, respectively and finally ${\omega _p} = {\left( {{n_{0e}}{e^2}/{m_e}{\varepsilon _0}} \right)^{1/2}}$ is the plasma frequency of conduction electrons. It is worth mentioning that the third term of the left side in Eq. (1) is the restoration force resulted from the displacement of electrons with respect to the positive background ions and parameter $\xi $ is used in order to modify the idealistic theory by the real experiments. In the simple  rigid body model where we suppose that all conduction electrons move identically, through some simple calculations, one can obtain $\xi  = 1/3$, however in the real case, its value depending on the MNP diameter is less than $1/3$ and vanishes by growing the particle size as one can expect for a bulk medium \cite{asef2020modified}. Here the electric field ${{\bf{E}}_n}$ includes the incident wave electric field ${{\bf{E}}_L}$ and the interactional term of $\sum\limits_j {{{\bf{E}}_{j \to n}}} $ where ${{\bf{E}}_{j \to n}}$ is the electric field at the place of $n$th MNP created by the electric dipole moment of the $j$th NP which after considering only the the fields of two nearest neighbors ($j = n \pm 1$), we can get
\begin{align}
\sum\limits_j {{{\bf{E}}_{j \to n}}}  = \frac{1}{{4\pi {\varepsilon _0}{d^3}}}\sum\limits_{j = n \pm 1}& {\left\{ {\left[ {{\mkern 1mu} 3{{{\bf{\hat e}}}_j}({{\bf{p}}_j}.{{{\bf{\hat e}}}_j}) - {{\bf{p}}_j}} \right]{\mkern 1mu} \left( {1 - ikd} \right){\mkern 1mu} } \right.} \nonumber \\&
\left. { + {k^2}{d^2}\left[ {{\mkern 1mu} {{\bf{p}}_j} - {{{\bf{\hat e}}}_j}({{\bf{p}}_j}.{{{\bf{\hat e}}}_j})} \right]} \right\},
\end{align}
where ${{\bf{\hat e}}_j}$ is a unit vector orientated from the $j$th MNP to the $n$th one, ${{\bf{p}}_{\bf{j}}} =  - Ze{{\bf{r}}_{\bf{j}}}$ is the electric dipole moment of $j$th MNP and $k$ is the wavenumber of the incident wave.

\section*{\label{sec:level3}Parallel Polarization}

Now, according to Fig. (1-a), we take the incident EMW electric filed parallel with the symmetry axis of MNPs array as following
\begin{align}
{{\bf{E}}_{L,\,{\rm{||}}}}({\bf{r}},t) = \frac{1}{2}E{e^{i(kx - \omega t)}}{{\bf{\hat e}}_z} + c.c.,
\end{align}
where $E$ and $\omega $ are the amplitude and frequency of the incident wave and $c.c.$ denotes the complex conjugate. \\
Considering Eqs. (2) and (3) in Eq. (1) implies that the oscillations of conduction electrons should be along the incident electric field and one can obtain the following equation for the displacement of center of mass of conduction electrons of the $n$th MNP
\begin{align}
{m_e}\frac{{{d^2}{z_n}}}{{d{t^2}}} &+ {m_e}\gamma \frac{{d{z_n}}}{{dt}} + {m_e}\xi \omega _p^2{z_n} = \frac{{ - eE}}{2}({e^{i(kx - \omega t)}} + c.c.) \nonumber\\ &+ \frac{{Z{e^2}}}{{2\pi {\varepsilon _0}}}\left( {\frac{1}{{{d^3}}} - \frac{{ik}}{{{d^2}}}} \right)\left( {{z_{n + 1}} + {z_{n - 1}}} \right).
\end{align}
Taking the solutions in the oscillational form of
\begin{align}
{z_n} = \frac{1}{2}{\tilde z_n}{e^{i(kx - \omega t)}}{{\bf{\hat e}}_z} + c.c.,
\end{align}
for the displacement of each MNP leads to the following recurrence relation for the amplitudes 
\begin{align}
{\tilde z_n} = {s_1}{\tilde z_{n + 1}} + {s_1}{\tilde z_{n - 1}} + {a_0},
\end{align}
Using the fundamental information from the discrete analysis course in mathematics, one can straightforwardly solve Eq. (6) and get
\begin{align}
{\tilde z_n} = {a_1}u_1^n + {a_2}u_2^n + {a_0},
\end{align}
where
\begin{align}
{u_{1,2}} = \frac{{1 \pm \sqrt {1 - 4s_1^2} }}{{2{s_1}}},\,\,\,\,\,{a_0} = \frac{{{z_0}}}{{1 - 2{s_1}}},
\end{align}
and also, we have the following definitions
\begin{align}
&{s_1} = \frac{{{E_d}}}{E},\,\,\,\,\,{E_d} = \frac{{ - Ze{z_0}}}{{2\pi {\varepsilon _0}}}\left( {\frac{1}{{{d^3}}} - \frac{{ik}}{{{d^2}}}} \right),\nonumber\\ &{z_0} = \frac{{eE}}{{{m_e}({\omega ^2} + i\omega \gamma  - \xi \omega _p^2)}},
\end{align}
constants ${a_1}$ and ${a_2}$ can be obtained from the boundary conditions as
\begin{align}
{a_1} = \frac{{\left[ {u_2^{N - 2}({u_2} - {s_1}) + ({s_1}{u_2} - 1)} \right]\,\left[ {({s_1} - 1){a_0} + {z_0}} \right]}}{{{u_1}u_2^{N - 2}({s_1}{u_1} - 1)({s_1} - {u_2}) + u_1^{N - 1}({s_1}{u_2} - 1)({u_1} - {s_1})}},
\end{align}
where ${N}$ is the number of NPs in the chain and ${a_1}$ converts to ${a_2}$ by interchanging ${u_1}$ to ${u_2}$ in Eq. (10). The details of mathematical operations for getting Eq.  (7) can be found in Ref. \cite{kheirandish2018polarization} and we do not bring it here, for brevity.\\
Considering the definition of electric dipole moment of $n$th MNP as ${p_n} =  - Ze{\tilde z_n}$ and the polarization vector as ${{\bf{\prod}}_n} = {n_{oe}}{{\bf{p}}_n} = {\varepsilon _0}{\chi _{n,{\rm{||}}}}{\bf{E_{n,{\rm{||}}}}}$, we derive the susceptibility as
\begin{align}
{\chi _{{\rm{n,}}\,\,{\rm{||}}}} = \left( {\frac{{{a_1}u_1^n + {a_2}u_2^n + {a_0}}}{{{z_0}}}} \right)\frac{{ - \omega _p^2}}{{({\omega ^2} + i\omega \gamma  - \xi \omega _p^2)}},
\end{align}
and consequently, the permittivity of $n$th MNP is achieved as
\begin{align}
{\left( {\frac{\varepsilon }{{{\varepsilon _0}}}} \right)_{n,\,\,{\rm{||}}}} = {\left( {n_{n,\,\,{\rm{||}}}^{(0)}} \right)^2} = 1 &+ {\chi _{{\rm{n,}}\,\,{\rm{||}}}}\, = 1 - \left( {\frac{{{a_1}u_1^n + {a_2}u_2^n + {a_0}}}{{{z_0}}}} \right)\nonumber\\ & \times \frac{{\omega _p^2}}{{({\omega ^2} + i\omega \gamma  - \xi \omega _p^2)}},
\end{align}
where $n_{n,\,\,{\rm{||}}}^{(0)}$ is the ordinary refractive index of $n$th MNP in the chain. Here, we considered only the dynamics of conduction electrons and in the Drude-Lorentz model, it is well-known that in order to take into account the role of bound electrons in the permittivity, one may rewrite Eq. (12) as 
\begin{align}
{\left( {\frac{\varepsilon }{{{\varepsilon _0}}}} \right)_{n,\,\,{\rm{||}}}}\, = {\varepsilon _\infty } - \left( {\frac{{{a_1}u_1^n + {a_2}u_2^n + {a_0}}}{{{z_0}}}} \right)\frac{{\omega _p^2}}{{({\omega ^2} + i\omega \gamma  - \xi \omega _p^2)}},
\end{align}
where, using experimental data of bulk medium, ${\varepsilon _\infty }$ is determined.\\
Simple calculations show that the parameter ${s_1}$ is very small than the unity at the entire interested frequency area in plasmonics and for the typical values of other physical variables, thus, from equation (8), by expanding ${u_{1,2}}$ and ${a_0}$ with respect to this small parameter and saving up to the third-order powers of ${s_1}$, one obtains the following
\begin{align}
&{a_0} \approx {z_0}\left( {1 + 2{s_1} + 4s_1^2 + 8s_1^3} \right),\,\,\,\,\,{u_1} \approx \left( {\frac{1}{{{s_1}}} - {s_1} - s_1^3} \right),\nonumber \\ &{u_2} \approx {s_1} + s_1^3.
\end{align}
Using the approximated Eqs. (14) in Eq. (10) and substituting it in Eq. (13) leads to the following simplified permittivity of particles located in different positions of the chain
\begin{align}
{\left( {\frac{\varepsilon }{{{\varepsilon _0}}}} \right)_{\,n,\,{\rm{||}}}} &= {\left( {\frac{\varepsilon }{{{\varepsilon _0}}}} \right)_{{\rm{single}}}} - \frac{{\omega _p^2\,}}{{({\omega ^2} + i\omega \gamma  - \xi \omega _p^2)}} \nonumber\\ &\times \left\{ \begin{array}{l}
{s_1} + 2s_1^2 + 3s_1^3,\,\,\,\,\,\,\,\,n = 1,\,N\\
2{s_1} + 3s_1^2 + 6s_1^3,\,\,\,\,\,n = 2,\,N - 1\\
2{s_1} + 4s_1^2 + 7s_1^3,\,\,\,\,\,n = 3,\,N - 2\\
2{s_1} + 4s_1^2 + 8s_1^3,\,\,\,\,\,4 \le n \le N - 3
\end{array} \right.,
\end{align}
where the first term is the relative permittivity of a single non-interactional NP \cite{asef2020modified}
\begin{align}
{\left( {\frac{\varepsilon }{{{\varepsilon _0}}}} \right)_{{\rm{single}}}} = {\varepsilon _\infty } - \frac{{\omega _p^2\,}}{{({\omega ^2} + i\omega \gamma  - \xi \omega _p^2)}},
\end{align}
and the second term is the contribution of interactions. Notice that there is a mirror symmetry in the system behavior and ${\left( {\varepsilon /{\varepsilon _0}} \right)_{\,N - 2,\,{\rm{||}}}},\,\,{\left( {\varepsilon /{\varepsilon _0}} \right)_{\,N - 1,\,{\rm{||}}}}$ and ${\left( {\varepsilon /{\varepsilon _0}} \right)_{\,N,\,{\rm{||}}}}$ are equal to ${\left( {\varepsilon /{\varepsilon _0}} \right)_{\,3,\,{\rm{||}}}},\,\,{\left( {\varepsilon /{\varepsilon _0}} \right)_{\,2,\,{\rm{||}}}}$ and ${\left( {\varepsilon /{\varepsilon _0}} \right)_{\,1,\,{\rm{||}}}}$, respectively. In the non-interactional limit when ${s_1} \to 0$, the permittivity approaches to the trivial solution of ${\left( {\varepsilon /{\varepsilon _0}} \right)_{\,n,\,{\rm{||}}}} = \,\,{\left( {\varepsilon /{\varepsilon _0}} \right)_{{\rm{single}}}}$. Doing more simplification and taking only the linear terms of small parameter ${s_1}$, equation (15) reduces to
\begin{align}
{\left( {\frac{\varepsilon }{{{\varepsilon _0}}}} \right)_{\,n,\,{\rm{||}}}} = {\left( {\frac{\varepsilon }{{{\varepsilon _0}}}} \right)_{{\rm{single}}}} - \frac{{\omega _p^2\,}}{{({\omega ^2} + i\omega \gamma  - \xi \omega _p^2)}} \nonumber\\ \times \left\{ \begin{array}{l}
{s_1},\,\,\,\,\,\,\,\,n = 1,\,N\\
2{s_1},\,\,\,\,\,2 \le n \le N - 1
\end{array} \right.,
\end{align}
in which the permittivity of the first and last particles reduces to the permittivity of MNP dimer obtained earlier in Ref. \cite{asef2020analytical} and for other particles the extra coefficient of $2$ shows the mutual interaction of each particle with two nearest neighbors.

\section*{\label{sec:level4}Perpendicular Polarization}

Now, according to Fig (1-b), we take the incident EMW electric filed perpendicular to the symmetry axis of MNPs chain as 
\begin{align}
{{\bf{E}}_{L,\, \bot }}({\bf{r}},t) = \frac{1}{2}E{e^{i(kx - \omega t)}}{{\bf{\hat e}}_y} + c.c..
\end{align}
Substituting Eq. (18) in Eq. (1), the linear motion equation of electrons of $n$th MNP in the chain is achieved as 	
\begin{align}
{m_e}\frac{{{d^2}{y_n}}}{{d{t^2}}} &+ {m_e}\gamma \frac{{d{y_n}}}{{dt}} + {m_e}\xi \omega _p^2{y_n} = \frac{{ - eE}}{2}({e^{i(kx - \omega t)}} + c.c.) \nonumber\\ &- \frac{{Z{e^2}}}{{4\pi {\varepsilon _0}}}\left( {\frac{1}{{{d^3}}} - \frac{{ik}}{{{d^2}}} - \frac{{{k^2}}}{d}} \right)\left( {{y_{n + 1}} + {y_{n - 1}}} \right).
\end{align}
By suggesting the solution in the form
\begin{align}
{y_n} = \frac{1}{2}{\tilde y_n}{e^{i(kx - \omega t)}} + c.c.,
\end{align}
and using the similar mathematical process which has been employed for the parallel polarization case of previous section, one can obtain the following equation for the amplitudes of the displacements
\begin{align}
{\tilde y_n} = {b_1}u_3^n + {b_2}u_4^n + {b_0},
\end{align}
where
\begin{align}
{u_{3,4}} = \frac{{1 \pm \sqrt {1 - 4s_2^2} }}{{2{s_2}}},\,\,\,\,\,{b_0} = \frac{{{y_0}}}{{1 - 2{s_2}}},
\end{align}
also, we have the following definitions
\begin{align}
{s_2} = \frac{{{{E'}_d}}}{E},\,\,\,\,\,{E'_d} = \frac{{Ze{y_0}}}{{4\pi {\varepsilon _0}}}\left( {\frac{1}{{{d^3}}} - \frac{{ik}}{{{d^2}}} - \frac{{{k^2}}}{d}} \right),\,\,\,\,\,{y_0} = {z_0},
\end{align}
finally, the constant ${b_1}$ obtained from the boundary conditions is
\begin{align}
{b_1} = \frac{{\left[ {u_4^{N - 2}({u_4} - {s_2}) + ({s_2}{u_4} - 1)} \right]\,\left[ {({s_2} - 1){b_0} + {y_0}} \right]}}{{{u_3}u_4^{N - 2}({s_2}{u_3} - 1)({s_2} - {u_4}) + u_3^{N - 1}({s_2}{u_4} - 1)({u_3} - {s_2})}},
\end{align}
and ${b_2}$ can be obtained from ${b_1}$ by interchanging ${u_3}$ to ${u_4}$ in Eq. (24).\\
After some simple and straightforward algebraic operations, one gets the following equation for the relative permittivity of the $n$th MNP in the chain
\begin{align}
{\left( {\frac{\varepsilon }{{{\varepsilon _0}}}} \right)_{n,\,\, \bot }}\, = {\varepsilon _\infty } - \left( {\frac{{{b_1}u_3^n + {b_2}u_4^n + {b_0}}}{{{y_0}}}} \right)\frac{{\omega _p^2}}{{({\omega ^2} + i\omega \gamma  - \xi \omega _p^2)}}.
\end{align}
Again, we can expand ${u_{3,4}}$ and ${b_{0,1,2}}$ with respect to the small parameter ${s_2}$, maintain up to its third order powers and get the following simplified relations for the relative permittivity
\begin{align}
{\left( {\frac{\varepsilon }{{{\varepsilon _0}}}} \right)_{\,n,\, \bot }} &= {\left( {\frac{\varepsilon }{{{\varepsilon _0}}}} \right)_{{\rm{single}}}} - \frac{{\omega _p^2\,}}{{({\omega ^2} + i\omega \gamma  - \xi \omega _p^2)}} \nonumber\\ &\times \left\{ \begin{array}{l}
{s_2} + 2s_2^2 + 3s_2^3,\,\,\,\,\,\,\,\,n = 1,\,N\\
2{s_2} + 3s_2^2 + 6s_2^3,\,\,\,\,\,n = 2,\,N - 1\\
2{s_2} + 4s_2^2 + 7s_2^3,\,\,\,\,\,n = 3,\,N - 2\\
2{s_2} + 4s_2^2 + 8s_2^3,\,\,\,\,\,4 \le n \le N - 3
\end{array} \right..
\end{align}
We see that the permittivity has the same form of the previous case of the parallel polarization and by changing ${s_1}$ to ${s_2}$ in Eq. (15), it converts to Eq. (26).

\begin{figure}
	\centerline{\includegraphics[scale=.45]{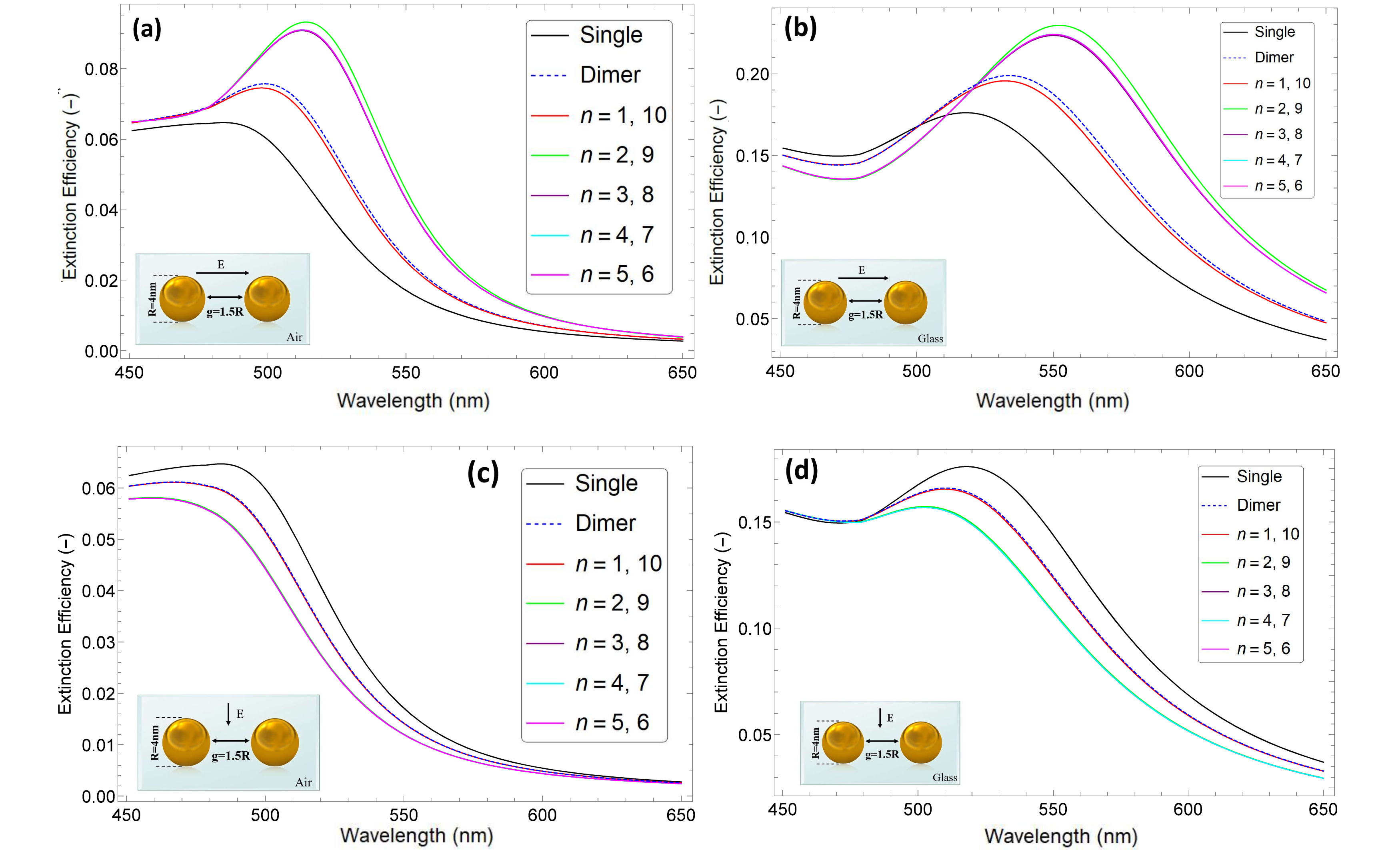}}
	\caption{Extinction efficiency spectra of gold NPs located at different positions, with diameter $R=4nm$ and separation gap $g=1.5R$, for two different states of polarization of (a), (b) parallel and (c), (d) perpendicular and for two different host media of (a), (c) Air and (b), (d) Glass.}
	\label{fig2}
\end{figure}

\begin{figure}
	\centerline{\includegraphics[scale=.45]{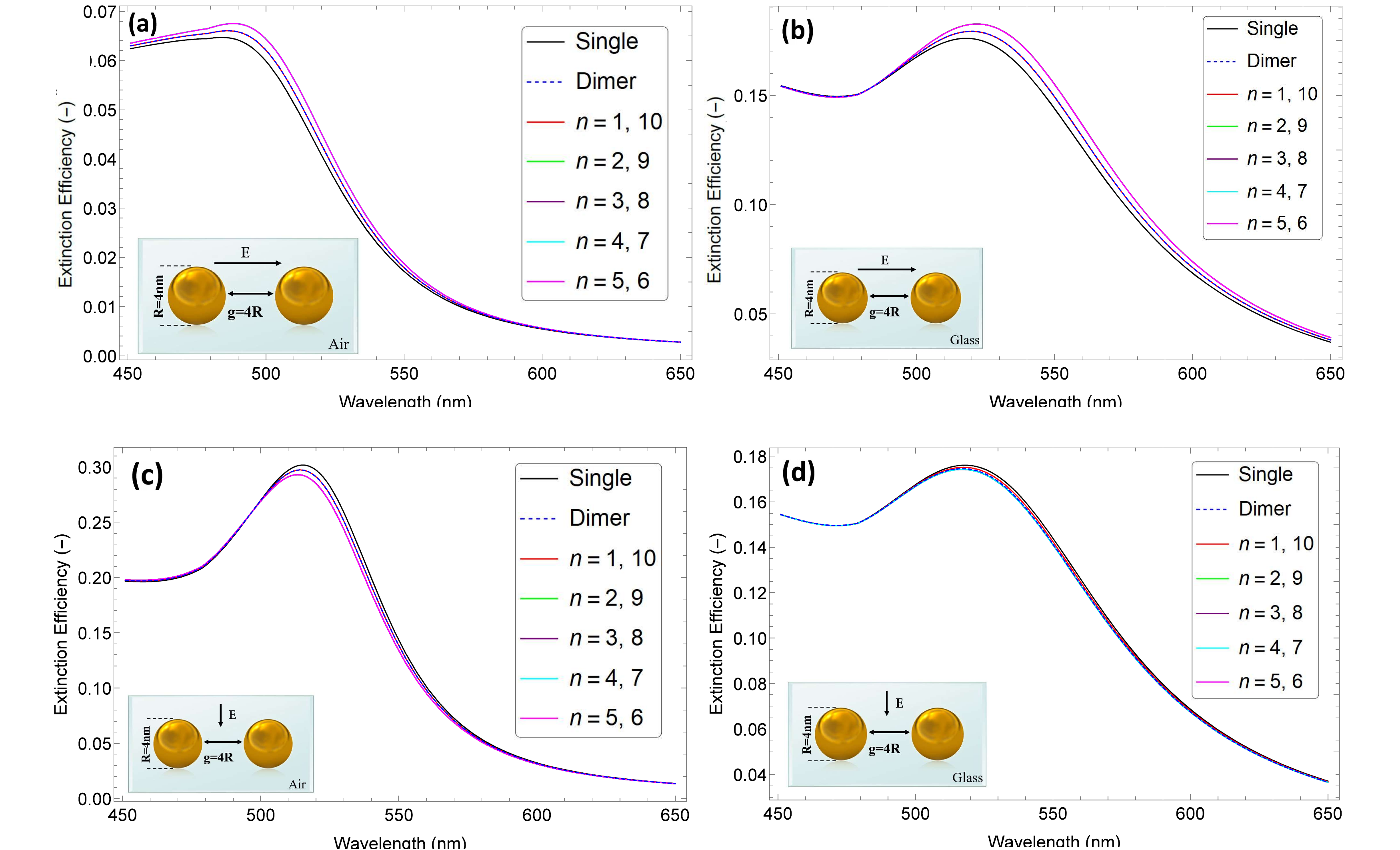}}
	\caption{Extinction efficiency spectra of gold NPs located at different positions, with diameter $R=4nm$ and separation gap $g=4R$, for two different states of polarization of (a), (b) parallel and (c), (d) perpendicular and for two different host media of (a), (c) Air and (b), (d) Glass.}
	\label{fig3}
\end{figure}

\section*{Extinction Cross Section}	

In practice, the permittivity of particles cannot be determined directly and the parameter which can be experimentally measured is the extinction cross section of NPs in a medium.  As a matter of fact, this parameter reflects the linear optical properties of NPs in the medium and it is defined as the sum of scattering and absorption cross sections according to the following equation resulted from the well-known Mie theory \cite{bohren2008absorption}
\begin{align}
{C_{ext}} = \frac{{2\pi }}{{{k^2}}}\sum\limits_{n = 1}^\infty  {(2n + 1){\mathop{\rm Re}\nolimits} ({a_n} + {b_n})} ,
\end{align}
\begin{align}
{a_n} = \frac{{m{\psi _n}(mx){{\psi '}_n}(x) - {\psi _n}(x){{\psi '}_n}(mx)}}{{m{\psi _n}(mx){{\xi '}_n}(x) - {\xi _n}(x){{\psi '}_n}(mx)}} ,
\end{align}
\begin{align}
{b_n} = \frac{{{\psi _n}(mx){{\psi '}_n}(x) - m{\psi _n}(x){{\psi '}_n}(mx)}}{{{\psi _n}(mx){{\xi '}_n}(x) - m{\xi _n}(x){{\psi '}_n}(mx)}} ,
\end{align}
where $x = 2\,\pi \,N\,a/\lambda $ is the size parameter, $m = {N_1}/N$ is the relative refractive index, where ${N_1}$ and ${N}$ are the refractive indices of particle and background medium, respectively, ${\psi _n}(x)$ and ${\xi _n}(x)$ are Riccati-Bessel functions.\\
In the limit of small-sized particles, i.e. $\,a/\lambda  <  < 1$, the extinction cross section reduces to
\begin{align}
{C_{ext}} &= 4x\pi {a^2}{\rm{Im}}\left\{ {\frac{{{m^2} - 1}}{{{m^2} + 2}}\left[ {1 + \frac{{{x^2}}}{{15}}\left( {\frac{{{m^2} - 1}}{{{m^2} + 2}}} \right)} \right.} \right.{\rm{ }}\nonumber\\& \times \left. {\left. {\frac{{{m^4} + 27{m^2} + 38}}{{2{m^2} + 3}}} \right]} \right\} + \frac{8}{3}{x^4}{\rm{Re}}\left[ {{{\left( {\frac{{{m^2} - 1}}{{{m^2} + 2}}} \right)}^2}} \right],
\end{align}
which in the case of very small particles or linear regime of x, reduces to the well-known equation of Rayleigh scattering as following  
\begin{align}
{C_{ext}} = 4x\pi {a^2}{\rm{Im}}\left( {\frac{{\varepsilon  - {\varepsilon _m}}}{{\varepsilon  + 2{\varepsilon _m}}}} \right).
\end{align}

\section*{Numerical Discussion} 

In order to study the effect of geometric and environmental parameters on the linear properties of interactional gold NPs, we numerically study the permittivity and extinction cross section of spherical gold NPs located in a linear chain. The effects of particles separation, kind of light polarization and medium in which the NPs are doped, are studied.  For all cases of future numerical analysis, let us consider a linear chain containing $N=10$ identical spherical gold NPs. For gold metal, the density of conduction electrons is ${n_{0e}} = 5.9 \times {10^{22}}c{m^{ - 3}}$. \\
In Figs. (2), for gold particles of $R=4nm$ diameter located at different positions in the chain, the extinction efficiency, i.e. the extinction cross section normalized by the cross section area of MNP ${C_{ext}}/\pi {a^2}$, is plotted for two different host media of air and glass and two different incident light polarizations when the gap distance between particles is $g = 1.5R$. It should be mentioned that the relation of extinction cross section ${C_{ext}}$ is used from Eq. (27), where we chose 10 terms of summation for calculations.  Figure (2-a) shows the variations of the extinction efficiency of all 10 NPs located at air with respect to the light wavelength for parallel polarization state. For comparison, we bring the extinction efficiency of single NP and dimer as well. As we proved earlier, for the permittivity of dimer, we can use the permittivity of special case of Eq. (17) for $n=1$, $N$ which is the reduced equation in the linear regime of ${s_1}$. Interaction of particles causes an increase in the extinction efficiency of all NPs of chain in comparison with the non-interactional single particle case. Also, interaction leads to the red-shift of plasmon resonance wavelength where the extinction experiences a maximum. The second and penultimate NPs, i.e. $n=2$ and $9$, have the largest extinction efficiency. The first and the last particles extinction is located at the middle of curves very close to the extinction of dimer that is their approximate values in the linear limit of ${s_1}$ which shows the validity of Eq. (17). Below the extinction curve of the second and penultimate NPs $n=2$ and $9$, the curves of other particles are located very close together however the sequence of their appearance from top to bottom is $n=(2,9) , (4,7), (5,6)$ and $(3,8)$. In Fig. (2-b), we study the effect of surrounded medium on the extinction efficiency by considering glass as the host medium. In this case, the damping factor related to the electron surface scattering changes and one can find more details in the Appendix. The dynamics of variations is the same but for all cases, the value of extinction efficiency increases approximately by the factor of $2$. Also, the places of plasmon resonance wavelength of all particles are shifted to large ones or in other words they experience red-shift in comparison with the air medium case, for example for the second and penultimate NPs $n=2$ and $9$ which have the largest extinction, the plasmon wavelengths are $513.8nm$ and $552.1nm$ for air and glass media, respectively. Figure (2-c) represents the variations of extinction efficiency for the case when the incident light electric field is perpendicular to the chain axis. In this case, the extinction efficiency of all particles decreases and its plasmon resonance wavelength experiences a blue-shift in comparison with the single particle case. The sequence of appearance of curves from top to bottom is as following $n= (1,10), (2, 9), (3, 8), (4, 7)$ and $(5, 6)$, however except particles $(1, 10)$ the curves related to the other particles are very close and practically coincide each other. In Fig. (2-d), with respect to the previous case, we only changed the host medium to glass. Here, like the parallel polarization, the dynamics of variations is the same of air medium case and for all cases, the value of extinction efficiency increases approximately by the factor of $2$. Also, the plasmon resonance wavelength places of all particles  experienced red-shift in comparison with the air medium case, for example for the first and last NPs $n=1$ and $10$ which have the largest extinction, the plasmon wavelengths are $482.8nm$ and $509.9nm$ for air and glass media, respectively. It is worth mentioning that the dynamics of displacing plasmon resonance wavelength and its amplitude as well changes with the same fashion of MNP dimer case which has been investigated earlier experimentally \cite{jain2007universal,jain2010plasmonic,rechberger2003optical} and theoretically \cite{asef2020analytical}.\\
\begin{figure}
	\centerline{\includegraphics[scale=.45]{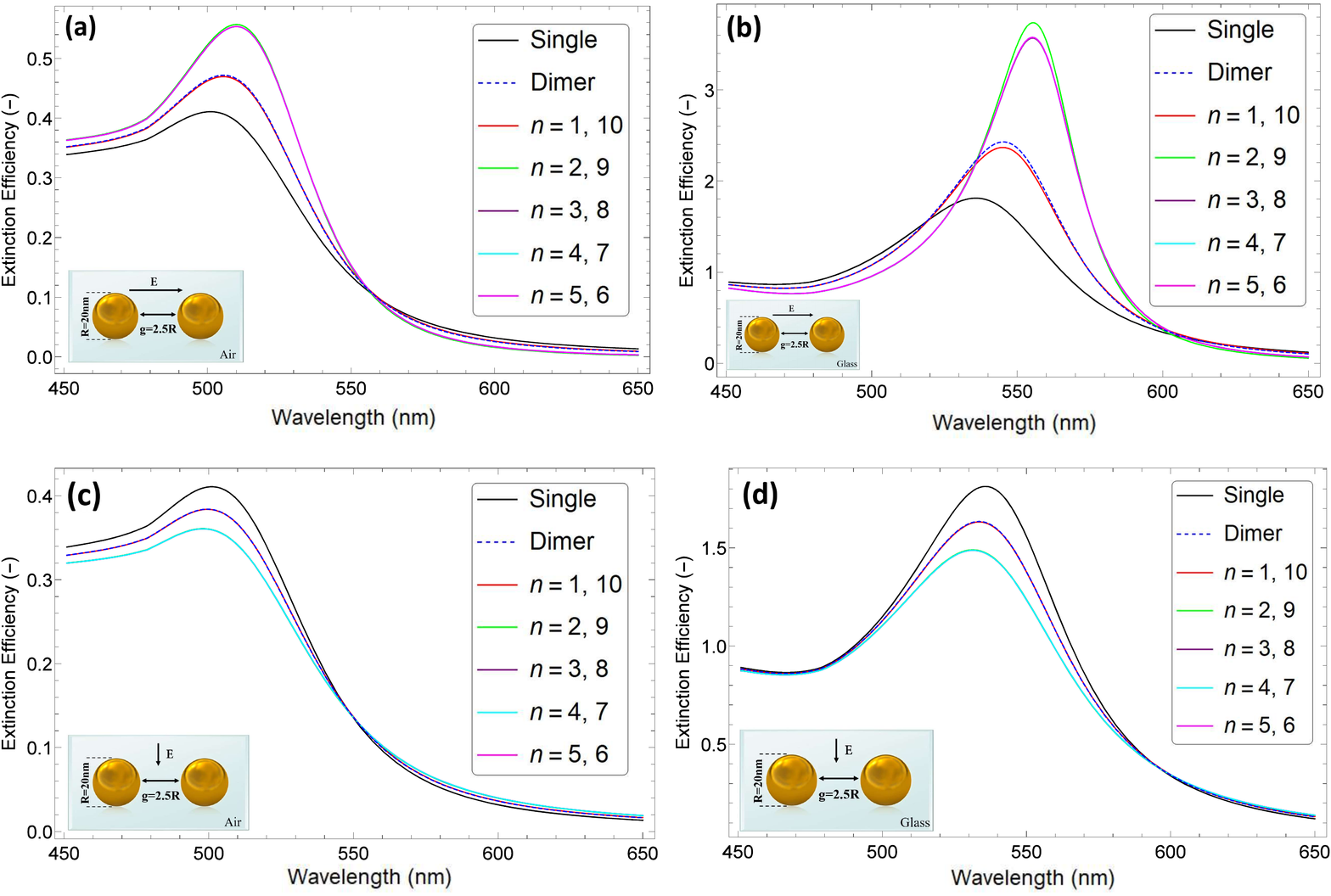}}
	\caption{Extinction efficiency spectra of gold NPs located at different positions, with diameter $R=20nm$ and separation gap $g=2.5R$, for two different states of polarization of (a), (b) parallel and (c), (d) perpendicular, for two different host media of (a), (c) Air and (b), (d) Glass.}
	\label{fig4}
\end{figure}

In Figs. (3), we increase the particle distances to $g=4R$ and plot the variations of extinction efficiency with respect to the wavelength for two different media and states of the incident light polarization. As one could expect, the an increase in the particles separation causes the weakening of dipolar interaction of particles and consequently, a decrease in the differences between single particle extinction efficiency and interactional particles of linear chain. However, the same behavior of previous curves of Figs. (2) is maintained here for all cases. For parallel state of polarization, the curves related to the extinction efficiency of the first and last NPs and dimer case are coincide to each other, practically the curves of other NPs of the chain totally overlap each other as well and the curve of single particle case is located in the bottom of other ones. For the perpendicular polarization, all curves of extinction efficiency related to the NPs of the chain, dimer case and single NP as well are very close together, however the same dynamics of close NPs spacing is maintained here.\\
\begin{figure}
	\centerline{\includegraphics[scale=.45]{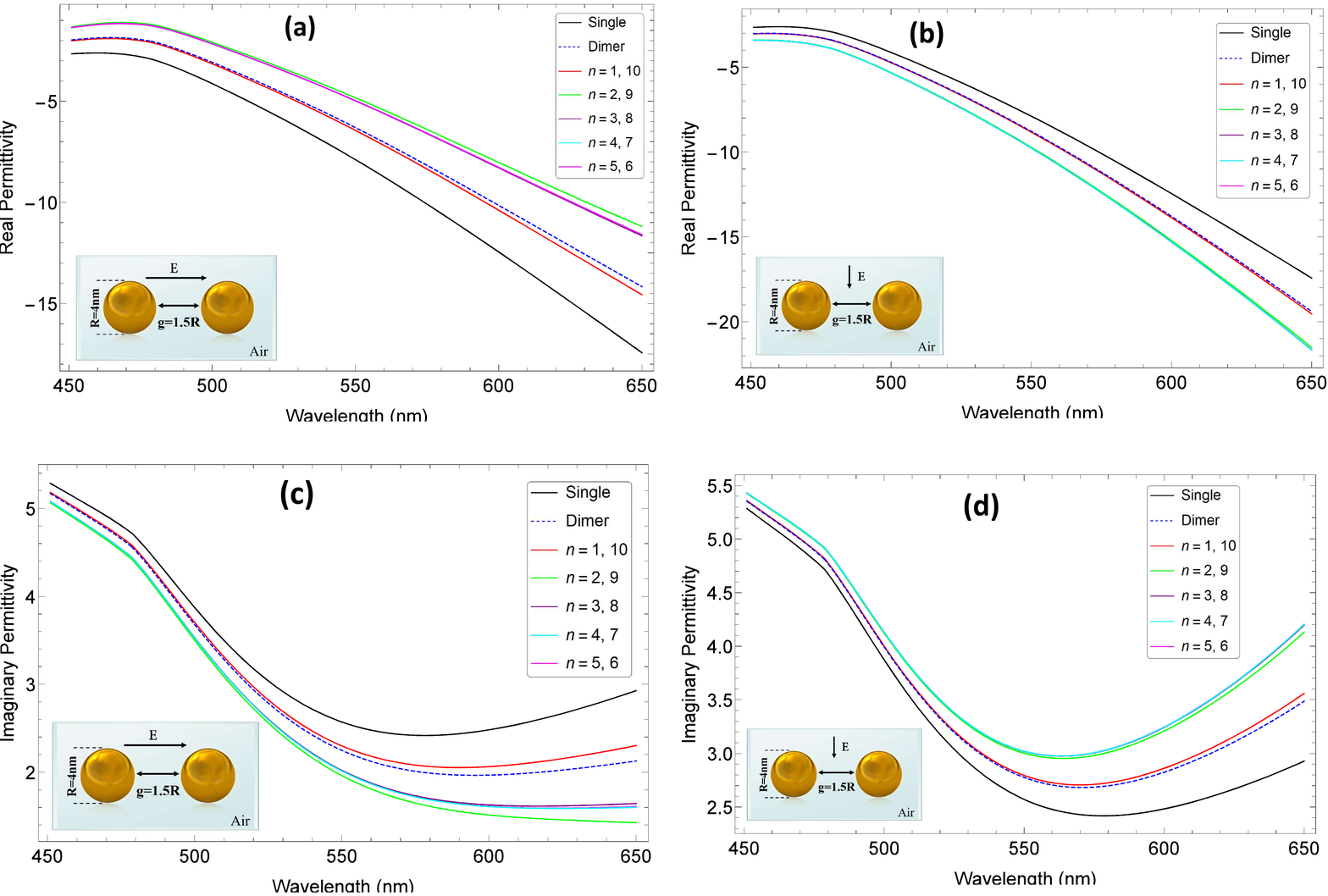}}
	\caption{Variations of (a), (b) real and (c), (d) imaginary parts of permittivity as a function of wavelength for gold NP located at different position with diameters of $R=4nm$ and interparticle separation gaps $g=1.5R$, for two different polarizations of (a), (c) parallel and (b), (d) perpendicular.}
	\label{fig5}
\end{figure}

In Figs. (4), we choose large NPs of diameter $20nm$ with spacing $g=2.5R$ and investigate the extinction efficiency of different NPs of the chain in different situations of previous figures. For all cases, the extinction efficiency increases and the plasmon resonance wavelength experiences red-shift for large NPs in comparison with similar cases of small NPs chain which is mainly related to the increase in the electron radiation damping. Near the plasmon resonance wavelength area, the same discipline of small NPs chain is governing, however in some cases, for large and small wavelengths, it is impaired. For example, in Fig. (4-a), which demonstrates the extinction efficiency of parallel polarization in the air, the regularity of location of different NPs curves changes after wavelength 560nm or in Fig. (4-b) which is related to the parallel polarization in glass medium, the regularity of curves changes not only for long wavelengths after $600nm$ but also for small wavelengths before $530nm$.\\
In Figs (5), we demonstrate the variations of the real and imaginary parts of permittivity of different NPs of chain with respect to the wavelength for diameter $R=4nm$, spacing $g=1.5R$, in air host medium and two different polarizations of incident light. For parallel case of polarization in Fig. (5-a), the real part of permittivity of different NPs is negative and it increases in comparison with the individual non-interactional single NP case. The order of appearance of curves with respect to the position of particles in the chain is as following $n=(2,9), (4, 7), (5, 6), (3, 8)$ and $(1, 10)$ from top to bottom, respectively. For perpendicular polarization case in Fig. (5-b), the real part of permittivity of different NPs in the chain is negative as well and it decreases in comparison with the permittivity of single NP. The sequence of location of curves from top to bottom is as following $n= (1, 10), (2, 9), (3, 8), (4, 7)$ and $(5, 6)$, however except particles $(1, 10)$ the curves related to the other particles are very close and approximately overlap totally each other. In Fig. (5-c), we show the variations of the permittivity imaginary part with respect to the wavelength for parallel polarization and air host medium. Interaction between NPs in the chain leads to the decrease in the imaginary part of permittivity and the order of appearance of different NPs is the same of real part but inversely from bottom to top. In Fig. (5-d), the permittivity imaginary part is plotted with respect to the wavelength for perpendicular polarization. Here, the dipole-dipole interaction of NPs causes an increases in the imaginary part of each NP of chain in comparison with the single particle case. The order of appearance of curves is the same of real part but inversely from bottom to top.\\
Figures (6-a)-(6-d) show the variations of the permittivity real and imaginary parts versus wavelength for large particles with diameter $R=20nm$ and spacing $g=2.5R$ located at air medium for two different kinds of polarization of incident light. With the same fashion of previous case of small particles, the permittivity changes with respect to the position of particles and kind of polarization. Additionally, here, an increase in the NP size leads to an increase in the real part of the permittivity and  a decrease in its imaginary part. As the imaginary part of permittivity is negative, in other words, an increase in the NPs size causes an increase in the absolute values of the real and imaginary parts of permittivity.
\begin{figure}
	\centerline{\includegraphics[scale=.45]{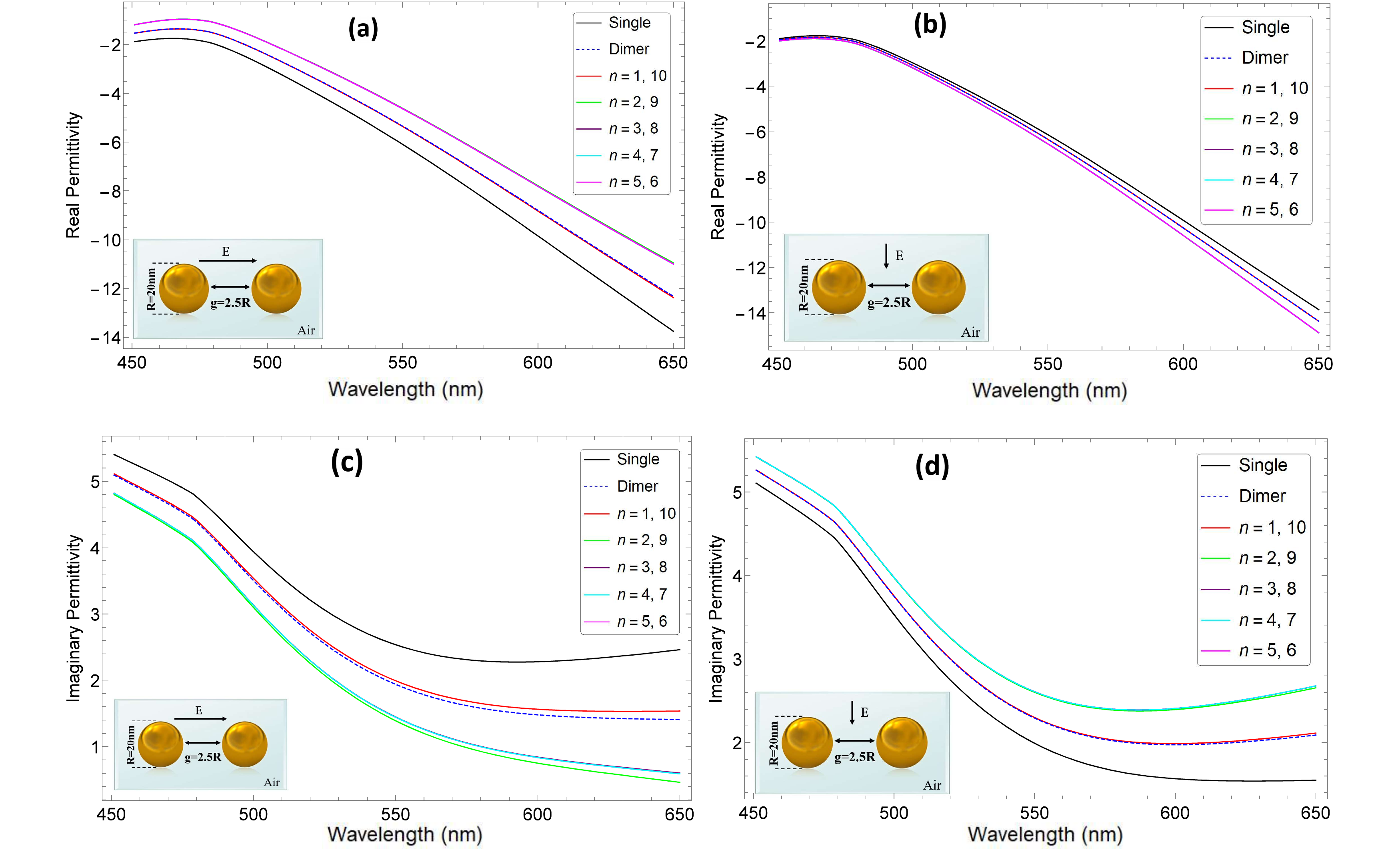}}
	\caption{Variations of (a), (b) real and (c), (d) imaginary parts of permittivity as a function of wavelength for gold NPs located at different positions with diameters of $R=20nm$ and interparticle separation gap $g=2.5R$, for two different states of polarization of (a), (c) parallel and (b), (d) perpendicular.}
	\label{fig6}
\end{figure}

\section*{Conclusions}

Based on a Drude-like model, the permittivity of interacting MNPs located in a linear periodic one-dimensional array has been derived analytically. The effects of interparticle spacing, size of particles, surrounding medium and state of incident light polarization on the extinction cross section and complex permittivity of each MNP have been investigated. It was found that for the parallel polarization of incident EMW, extinction efficiency increased and the SPR wavelength red-shifted while for the perpendicular polarization, interaction of particles caused a decrease in both quantities of extinction efficiency and SPR wavelength. The most variations of the extinction efficiency and displacement in the SPR wavelength belonged to the second and penultimate particles whereas the first and last particles experienced the least variations with respect to the single particle case. Finally, the impact of the interaction on the real and imaginary parts of the permittivity was studied and it was noticed that for the parallel polarization, mutual dipolar interaction among particles resulted in a decrease in the absolute value of the real and imaginary parts of the complex permittivity of each MNP in comparison with the individual particle case while they increased for the perpendicular state of polarization. \\

\begin{figure}
	\centerline{\includegraphics[scale=.45]{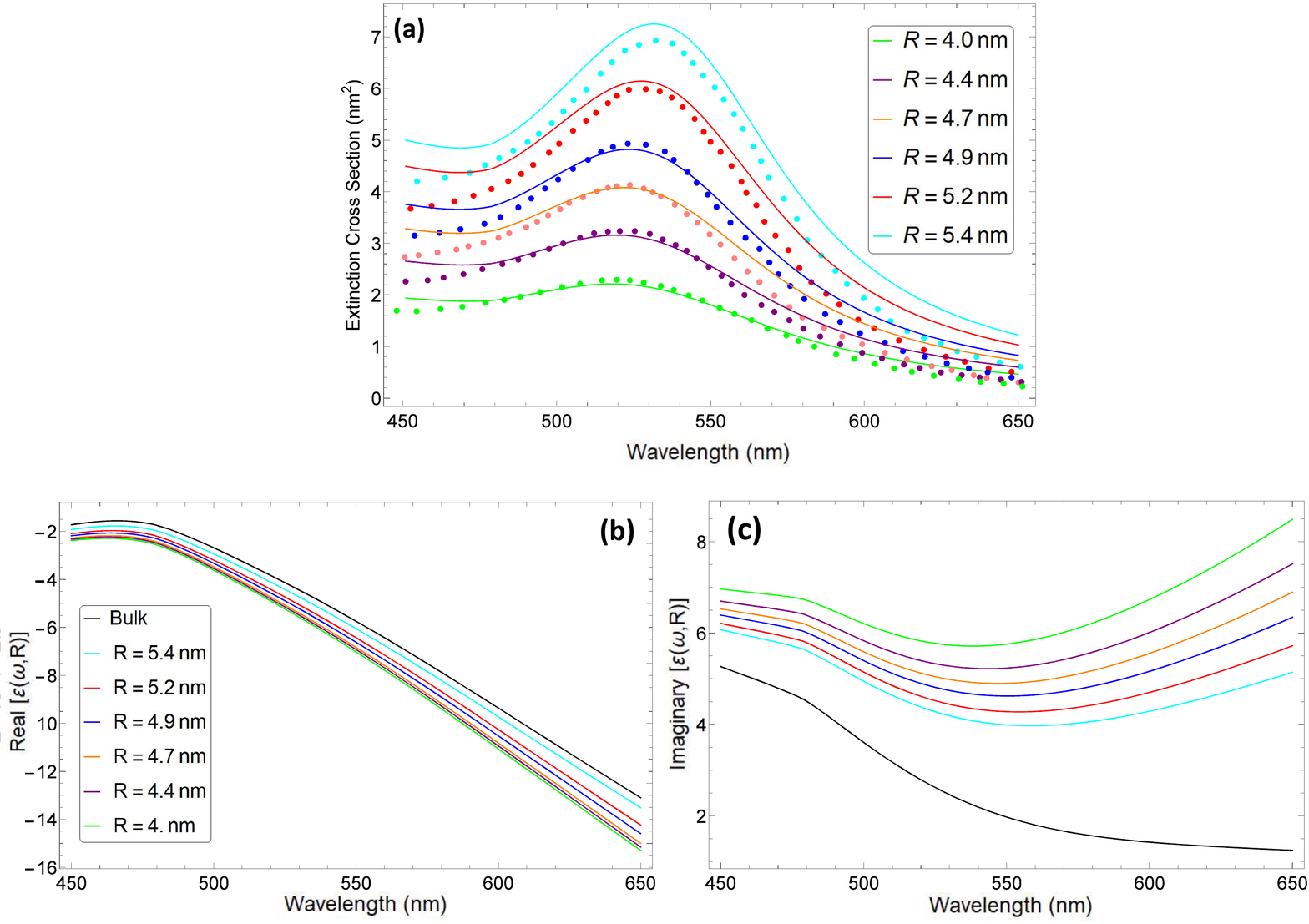}}
	\caption{(a) The calculated extinction cross sections (solid lines) and the experimentally measured ones (dotted lines), (b) real and (c) imaginary parts of the permittivity in dependence of wavelength for NPs diameters of 4, 4.4, 4.7, 4.9, 5.2, 5.4nm (from bottom to top).}
	\label{fig7}
\end{figure}
\begin{figure}
	\centerline{\includegraphics[scale=.45]{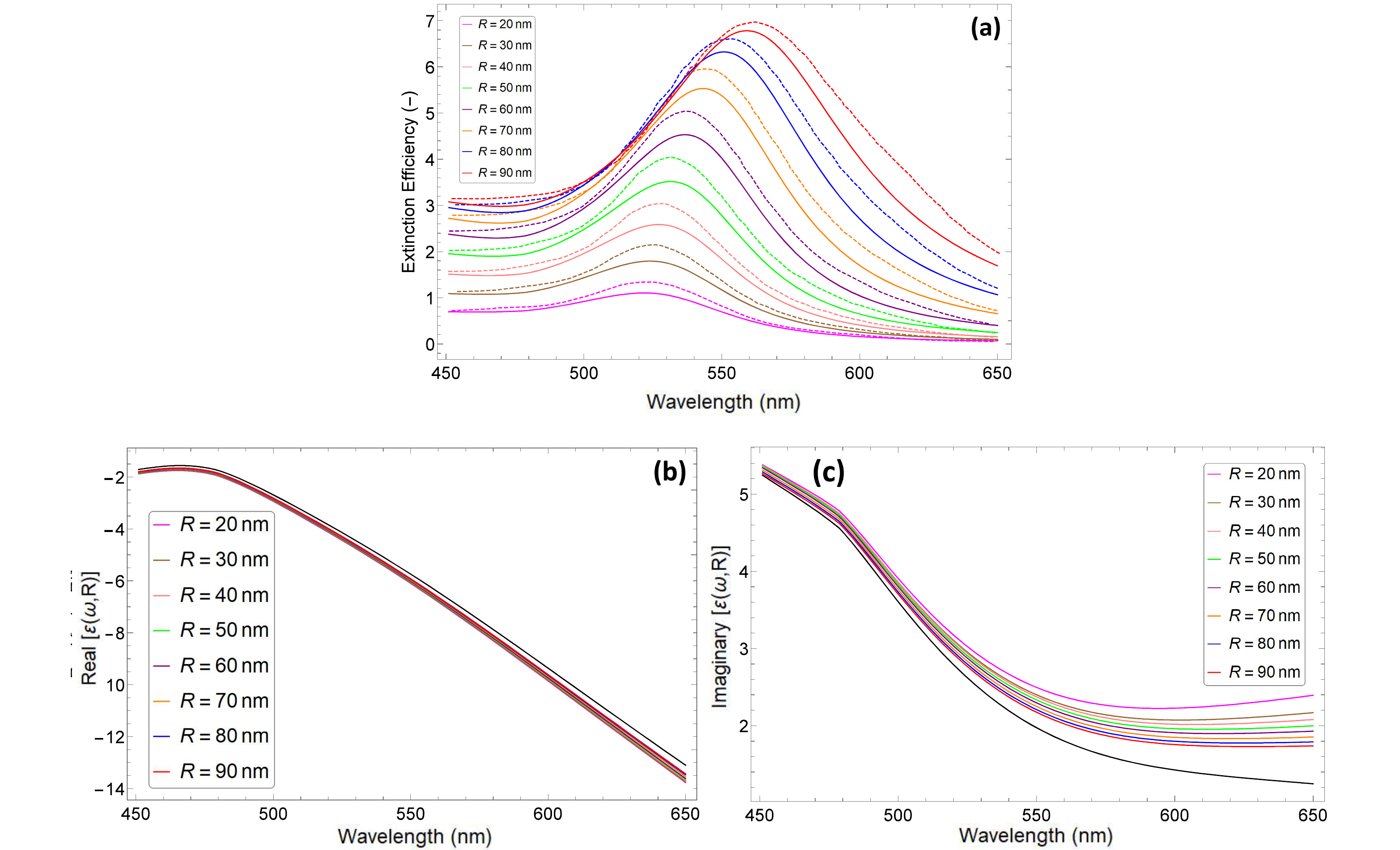}}
	\caption{(a) The calculated extinction efficiency (solid lines) and the experimentally measured ones (dotted lines), (b) the real and (c) the imaginary parts of permittivity in dependence of wavelength for NPs diameters of 10, 20, 30, 40, 50, 60, 70, 80 and 90nm (from bottom to top).}
	\label{fig8}
\end{figure}

\section*{Appendix}
In this appendix, using experimental data of extinction cross section of individual NPs, we determine the permittivity of a single NP by obtaining free phenomenological parameters of $\xi $ and $\gamma $. The extinction cross section of small gold NPs ranged from 4nm to 5.4nm \cite{kreibig2013optical} and large gold NPs ranged from 20nm to 90nm \cite{starowicz2018tuning,hong2013optimal} can be found in experimental investigations. For small NPs in Ref\cite{kreibig2013optical}, the sample is prepared by ion exchange method via heating metal films onto a Na-Glass surface, when the diffusion of the metal ions into the matrix occurs at elevated temperatures by $N{a^ + } - A{u^ + }$ ion exchange. For large particles in Ref\cite{hong2013optimal}, gold NPs synthesized by multiple reduction process, where the reduction of $HAuC{l_4}$ with trisodium citrate is used in aqueous solution, in the first step for obtaining NPs with average size of 17nm. Next, $HAuC{l_4}$ and Hydroxylamine were added into the 17nm gold seed solution to obtain NPs with larger size of 30nm. Repeating this process with appropriate concentration of $HAuC{l_4}$, Hydroxylamine and gold NPs, can produce larger sizes of NPs. During a trial and error process, the value of the free parameters $\xi $ and $\gamma $ in the permittivity of an NP with a given diameter are extracted so that by using this permittivity in the extinction cross section of Eq. (27), good approvement reveals between experimental and theoretical data. For small NPs, in Ref. \cite{asef2020modified}, we have already determined these parameters as a function of wavelength and NP radius as following
\begin{align}
\xi  = {\xi _0} + {c_1}\frac{1}{a} + {c_2}\frac{1}{{{a^2}}} + {c_3}\frac{1}{{{a^3}}},
\end{align}
where 
\begin{align}
{\xi _0} &=  - 2.75 \times {10^{ - 1}},\,\,\,{c_1} = 1.75nm,\,\,\,{c_2} =  - 3.66n{m^2},\nonumber\\{c_3} &= 2.57n{m^2},
\end{align}
and $a$ is in nm. For the damping factor of NP, we use the following expression
\begin{align}
\gamma  = {\gamma _{{\rm{e}} - {\rm{e}}}} + {\gamma _{{\rm{e}} - {\rm{ph}}}} + {\gamma _{{\rm{rad}}}} + {\gamma _{{\rm{surf}}}} + {\gamma _{{\rm{cor}}}},
\end{align}
where ${\gamma _{{\rm{e}} - {\rm{e}}}},\,\,{\gamma _{{\rm{e}} - {\rm{ph}}}},\,\,{\gamma _{{\rm{rad}}}}$ and $\,\,{\gamma _{{\rm{surf}}}}$ are the damping factors\cite{asef2020modified} related to the electron-electron scattering\cite{lawrence1976electron,lawrence1973electron}, the electron-phonon scattering\cite{holstein1954optical,holstein1964theory}, the radiation\cite{liu2009reduced} and the scattering of electrons by the NP surface\cite{genzel1975dielectric,coronado2003surface,liu2004synthesis,berciaud2005observation}, respectively. Also, for ${\gamma _{cor}}$ we used the following equation
\begin{align}
{\gamma _{cor}} \times {10^{ - 20}} = {d_1}\frac{1}{{{\lambda ^2}}} + {d_2}\frac{{{a^2}}}{{{\lambda ^2}}} + {d_3}\frac{{{a^3}}}{{{\lambda ^2}}},
\end{align}
where 
\begin{align}
{d_1} &=  - 2.17n{m^2},\,\,\,{d_2} = 7.64 \times {10^{ - 3}},\nonumber\\{d_3} &=  - 1.25 \times {10^{ - 2}}n{m^{ - 1}},\,
\end{align}
and all lengths are in nm.
Figure (7-a) shows the experimental\cite{kreibig2013optical} and theoretical data for the extinction cross section of gold NPs doped in the glass with diameters ranged from 4nm to 5.4nm. According to our model, the real and imaginary parts of permittivity are plotted in Figs. (7-b) and (7-c), respectively.\\
For large NPs, using the data of Ref. \cite{starowicz2018tuning,hong2013optimal}, we propose the following values for the coefficients of Eqs. (32) and (35):
\begin{align}
{\xi _0} &=  - 6.19 \times {10^{ - 4}},\,\,\,{c_1} = 8.23 \times {10^{ - 2}}nm,\nonumber\\{c_2} &=  - 1.13n{m^2},\,\,\,\,{c_3} = 5.19n{m^2},\,
\end{align}
and 
\begin{align}
{d_1} &=  - 2.08n{m^2},\,\,\,{d_2} = 4.38 \times {10^{ - 4}},\nonumber\\{d_3} &=  - 6.34 \times {10^{ - 5}}n{m^{ - 1}}.
\end{align}
Figure (8-a) demonstrates the variations of extinction efficiency as a function of wavelength for different sizes of large NPs suspended in a water medium. Experimental data are used from Ref.\cite{starowicz2018tuning,hong2013optimal} and the theoretical curves are obtained from our model. In Figs. (8-b) and (8-c), the real and imaginary parts of permittivity are plotted as a function of wavelength.\\\\

\section*{Competing interests}

The authors declare no competing interests.

\section*{Author contributions}

This work is a part of study of A. K. PhD dissertation under supervision of H. M. and N. S. J. The main idea is conceived by N. S. J and carried out by A. K. Calculations were checked by H. M. The draft is written by A. K. and N. S. J. All authors discussed the results and made comments on the manuscript.

\section*{Data Availability}
The data that support the findings of this study are available from the corresponding author upon reasonable request.

\bibliography{sample}
 
\end{document}